\documentclass[a4,twoside,showpacs]{revtex4}
\usepackage{mathptmx}
\usepackage{graphicx}
\usepackage{amssymb}

\usepackage{fancyhdr}
\begin{document}

\bibliographystyle{apsrev}

\title{Optimizing Nuclear Reaction Analysis (NRA) using Bayesian Experimental Design}
\author{U. von Toussaint, T. Schwarz-Selinger, and S. Gori}
\affiliation{Max-Planck-Institut f\"ur Plasmaphysik, EURATOM Association,
Boltzmannstr. 2, 85748 Garching, Germany}

\begin{abstract}
Nuclear Reaction Analysis with ${}^{3}$He holds the promise to measure Deuterium depth profiles up to large depths.
However, the extraction of the depth profile from the measured data is an ill-posed inversion problem. Here we demonstrate how Bayesian Experimental
Design can be used to optimize the number of measurements as well as the measurement energies to maximize the information gain.
Comparison of the inversion properties of the optimized design with standard settings reveals
huge possible gains. Application of the posterior sampling method allows to optimize the experimental settings interactively during the measurement process.
\rule{14 cm}{0.4 pt}
\end{abstract}

\maketitle
\section{Introduction}
The rising price for oil has recently shifted the focus to other possible sources of energy, preferably without
adverse effects to the environment. One of the methods presently being developed is nuclear magnetic fusion. The objective of fusion research is to harness the energy provided by the fusion of hydrogen isotopes. In the fusion experiment ITER, presently under construction in Cadarache, France, the necessary data to design and operate anelectricity-producing plant shall be gained. ITER is a tokamak, an intermittent operating device in which strong magnetic fields confine a torus-shaped plasma.
\begin{figure}
\includegraphics[width = 3 cm]{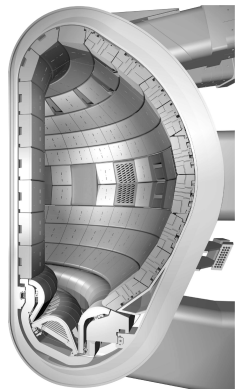}
\caption{\it In-vessel view of ITER. The surface of the main chamber is Beryllium, the material of the divertor (in lower part of the vacuum vessel)
is carbon and tungsten. Source:\cite{Iter08}, published with permission of ITER.}
\label{fig_iter}
\end{figure}
Since the confinement is not perfect (and must not be) there are always interactions between the plasma and the plasma-facing (wall) components (PFCs) which have to be taken into account.
One of the key aspects in the licensing process of ITER is a strict upper limit of the total amount of radioactive tritium accumulated in the vessel walls, which is presently at 700g tritium\cite{Iter08}.
The prediction of the amount of retained tritium is complicated by the material choice of ITER (Fig.\ref{fig_iter}): The main vessel walls are Beryllium, the strike-points are made of carbon (CFC) and the other parts of the divertor are tungsten. During the operation of ITER the interaction of the plasma and high energy 14MeV-neutrons with the vessel walls
will lead to erosion, redeposition, material mixing and alloy formation.
Since even the hydrogen retention properties of pure materials are still subject to current research,
a significant amount of additional experimental data is required to develop and calibrate the theoretical models which will be needed to process the
huge number of material combinations created in ITER.

However, even the first step - measuring hydrogen depth profiles in material composites - is challenging for many reasons; here we will mention only two: a) Hydrogen and its isotopes are very volatile, which can easily distort measurements of depth profiles and b) Hydrogen is usually the main component of the residual gas in vacuum chambers which precludes the use of many well-established analysis methods.

One method which holds great promise to overcome these difficulties is the Nuclear Reaction Analysis (NRA) of deuterium using ${}^{3}$He as probing particle.
It is a specific and sensitive
method, and has a sufficient analysis depth. However every data point takes about 30min to measure and the extraction of the concentration depth profile is an ill-posedinversion problem
requiring the deconvolution of the measured data vector, here even more challenging than in Rutherford Backscattering\cite{Toussaint2000}.
Therefore the experimental setup (ie the choice of the analysis energies) should provide a maximum of information.
So far the most common choice of the ${}^{3}$He energies for the measurements was simply equidistant.
Using Bayesian Experimental Design the performance of the method can be improved considerably (in some cases up to orders of magnitude) and quantitative measures can bederived
about the expected utility of further (time consuming) measurements.
section{Nuclear Reaction Analysis}
The basic principle of Nuclear Reaction Analysis is straightforward:
The sample is subjected to an energetic ion beam (here ${}^{3}$He) with initial energy ${E}_{i}^{0}$ and incoming angle $\phi$, which reacts predominately with the species of interest (Deuterium)
and the products of the reaction are measured under a specified angle $\theta$. Given the total number of impinging ions N${}_{i}$, the energy dependent cross-section of the reaction $\sigma\left(E\right)$,
the efficency of the detection and the geometry of the set-up $\mu$ the measured total signal counts $d_{i}$ depend (in the limit of small concentrations) linearly on the concentration profile $c\left(x\right)$ of the species in the depth $x$:
\begin{equation}
d_{i}=d\left(E_{i}^{0}\right)=\mu \mathrm{N}_{i}\int_0^{x\left(E_{i}^{0}\right)}\!dx\, \sigma\left(E\right)c\left(x\right)= \mu \mathrm{N}_{i}\int_0^{x\left(E_{i}^{0}\right)}\!dx\, \sigma\left(E\left(x,E_{i}^{0}\right)\right)c\left(x\right) + \epsilon_{i},
\label{equation_1}
\end{equation}
where $\epsilon_{i}\sim{\phantom a}N\left(0,\sigma_{i}\right)$ represents normal distributed noise.
Repeated measurements with different initial energies of ${}^{3}$He provide increasing information
about the Deuterium depth profile.
The question addressed in the following is: {\it Given a set of already measured data $d\left(E_{i}^{0}\right)$ which measurement energy should be chosen next?}

To evaluate Eq. \ref{equation_1} we first need to specify the cross section $\sigma\left(E\right)$ and the energy $E\left(x\right)$ of the
incident particle on its path through the sample.
\subsection{Cross-Section}
The relevant cross-section for the reaction D+${}^{3}\mathrm{He}\rightarrow p+{}^{4}\mathrm{He}$+ 18.352 MeV (in standard notation written as $\mathrm{D}\left({}^{3}\mathrm{He},p\right){}^{4}\mathrm{He}$) has been (re-)measured recently \cite{Alimov05} in the range of
550 keV to 6MeV and the obtained cross-section values have been given in tabular form. Using the same method as \cite{Alimov05} we added several cross-section measurements
at energies below 690keV and fitted both data sets taking into account also earlier measurements \cite{Moeller80,Bosch92}
\begin{equation}
\sigma\left(E\left[MeV\right]\right)=829.98*\frac{E^{2.83962}\left(0.270713*e^{-2.2158E}+0.0182765\right)}{E^{3.47626}+0.270713*e^{-1.17229E}-0.00123669}\left[mb\right].
\end{equation}
using the Levenberg-Marquardt algorithm minimizing the $\chi^{2}$-misfit with the variance set to $d_{i}$.
\begin{figure}
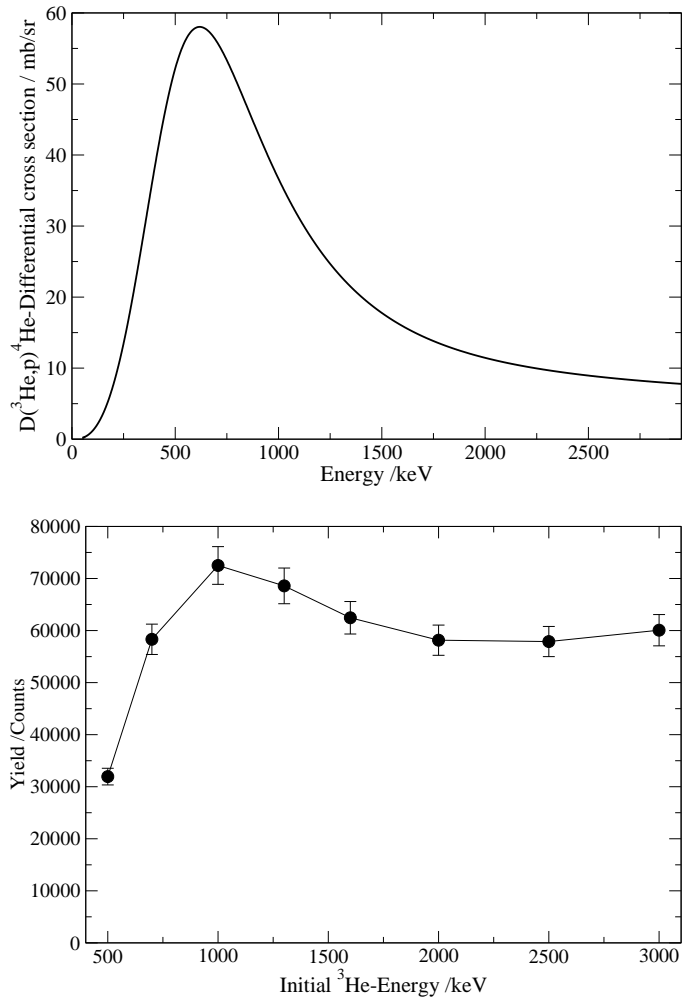

\begin{minipage}[t]{0.50\textwidth}
\includegraphics[angle=0,width=1.0\linewidth]{cross_section.eps}
\label{fig_yield_a}
\end{minipage}
\hfil
\begin{minipage}[t]{0.50\textwidth}
\includegraphics[angle=0,width=1.0\linewidth]{yield2.eps}
\label{fig_yield_b}
\end{minipage}
\caption{a) Differential cross-section of the nuclear reaction D(${}^{3}$He,p)${}^{4}$He in the laboratory system with a reaction energy of Q=18.352 MeV (left).
b) On the right hand side the typical results for an NRA measurement are shown (simulated data of a tungsten sample with an exponentially decaying D concentration profile (cf. Eq.\ref{eq_mock_data}).
The uncertainties due to the counting statistics are usually dominated by the uncertainty of the analysis current.}
\label{fig_yield}
\end{figure}
The cross-section is plotted in Fig.\ref{fig_yield}a and shows a broad maximum around 630 keV and is above 3 MeV nearly constant at 8 mb/sr up to 6 MeV (above that there are no data available).
The reaction energy is very high (Q=18.352 MeV) and most of the energy is transferred to the resulting proton. This leads to a very good S/N-ratio of the measurement
because other particles can easily be separated by energy.
\subsection{Energy Loss}
The energy loss of the impinging ${}^{3}$He-ion in the sample is determined by the {\it{stopping power}} $S(E)$ of the sample
\begin{equation}
\frac{dE}{dx}=-S(E),
\end{equation}
which can be solved to get the depth dependent energy $E_{i}(x)$ for different initial energies ${E}_{i}^{0}$. Parameterizations and tables of $S$ for different elements are given in \cite{Tesmer95}.
Since the amount of hydrogen in the sample is usually well below $1\%$ (with the exception of a very thin surface layer), the
influence of the hydrogen concentration on the stopping power can be neglected in most cases.

\subsection{Simulation of Mock Data}
To simulate mock data for typical accelerator parameters a tungsten sample $\left(\rho = 19.3\mathrm{g/cm}^{3}\right)$ with a (high) surface concentration of 12$\%$ Deuterium,
followed by an exponentially decaying Deuterium concentration
down to a constant background level, described by
\begin{equation}
c(x)=a_{0}*\exp\left(-\frac{x}{a_{1}}\right)+a_{2}=0.1\exp\left(-\frac{x}{2.5*10^{18}\frac{\mathrm{at}}{\mathrm{cm}^2}}\right)+0.02
\label{eq_mock_data}
\end{equation}
has been used \footnote{Generally ion-beam analysis methods are sensitive to the areal density of the species $\left(\mathrm{at/cm^{2}}\right)$ which can be converted into a depth scale if the density of the material is known. In tungsten $a_{1}=2.5*10^{18}\frac{\mathrm{at}}{\mathrm{cm}^2}$ correspond to $a_{1}=$395nm.}.
The corresponding mock data for a set of initial energies E${}^{0}$=\{500, 700, 1000, 1300, 1600, 2000, 2500, 3000\}keV is shown in Fig. 2b.
The variations in the detected yields display the interplay of the increasing range of the ions with increasing energy and the reduced cross-section at higher energies modulated with
the decreasing Deuterium concentration at larger depths. The increase of the signal by raising the initial energy from 2500 keV to 3000 keV is already caused by the constant Deuterium background of 2\%.
The accelerator time which would be needed to obtain the 8 data points is around one working day taking into account the necessary interleaved calibration measurements:The ion bombardment causes an energy and depth dependent loss of Deuterium.
Commonly a first order correction is applied by normalizing the yields with respect to the yields obtained
from repeated calibration measurements using the same (typically low, e.g. 690keV) initial energy\footnote{Remark: This approach of taking into account the Deuterium loss, although in widespread use, will almost always introduce a systematic bias in the derived concentration profile:
 The loss of Deuterium near the surface is used to correct the signal resulting from the overall Deuterium concentration profile, where the losses are usually different.}.

The uncertainty of the detector is given by Poisson-statistics. However, fluctuations in the beam current measurements are very often the dominating factor, affecting the pre-factor
$\mathrm{N}_{i}$ in Eq.\ref{equation_1}. An accuracy of up to 3$\%$ can be achieved (e.g. by using the number of Rutherford-scattered ${}^{3}$He ions on a thin gold-coating on top of
the sample as reference). The error of the renormalization procedure is harder to quantify.
For simplicity we will use $\sigma_{i}=\mathrm{max}\left(5\%d_{i},\sqrt{d_{i}}\right)$ as uncertainty of the data in the following, acknowledging that there is room forimprovement.
\section{Bayesian Experimental Design}
Bayesian Experimental Design (BED) offers the tempting possibility to actively select (and optimize) the experimental
parameters for the next measurement(s) based on objective criteria. Especially if measurements are expensive or time consuming (like in the case of
energy changes of an accelerator) it is a huge advantage to know where to look next, so as to learn as much as possible.
The problem of experimental design has already been studied by Lindley back in 1956 \cite{Lindley56} in a Bayesian setting
and Fedorov published his influential book in 1972 \cite{Fedorov72} - but the limitations in computational power limited
the application of experimental design almost always to simple (linear) problems. This situation changed in the recent
years and consequently there is a renewed interest to apply BED also to (non-linear) real-world problems (see e.g. \cite{Loredo03} and references therein or e.g. \cite{Preuss08,Dreier07,Fischer04,Caticha00}. Not surprising the interest is biggest in branches of physics where the experimental possibilities are severely restricted: Astronomy, Fusion research,...
Given the excellent account of BED in \cite{Loredo03} we only summarize the key principles:
In a first step an appropriate {\it utility function U} has to be agreed upon: It describes the value which we
assign to the measurement results of an experiment and may include parameters like costs of an experiment, duration,
parameter uncertainty etc. Several utility functions are considered in \cite{Chaloner95}.
With focus on parameter estimation it was proposed \cite{Lindley56, Bernardo79} to use the Kullback-Leibler divergence (KLd) between the posterior and the prior distributions as utility function. The KLd for a new datum D is given by
\begin{equation}
U_{KL}\left(D,\underline{d},\eta\right)=\int\!d\underline{\alpha}\; p\left(\underline{\alpha}|D,\underline{d},\eta\right)\log\left[\frac{p\left(\underline{\alpha}|D,\underline{d},\eta\right)}{p\left(\underline{\alpha}|\underline{d},\eta\right)}\right].
\label{eq_utility}
\end{equation}
Next we try to identify the action $\eta$ which maximizes the expected utility. 'Expected' utility because we have to account for the prediction uncertainty
for $D$.
To compute the expected utility (EU) we have to average over the new datum D weighted by the marginal likelihood for the new datum
given the observation of the old data $\underline{d}$
\begin{eqnarray}
EU\left(\underline{d},\eta\right)&=&\int\!dD\;p\left(D|\underline{d},\eta\right)\cdot U_{KL}\left(D,\underline{d},\eta\right)\nonumber\\
&=&\int\!dD\;p\left(D|\underline{d},\eta\right)\int\!d\underline{\alpha}\; p\left(\underline{\alpha}|D,\underline{d},\eta\right)
\log\left[\frac{p\left(\underline{\alpha}|D,\underline{d},\eta\right)}{p\left(\underline{\alpha}|\underline{d},\eta\right)}\right]\nonumber\\
&=&\int\!dD\;p\left(D|\underline{d},\eta\right)\int\!d\underline{\alpha}\;\frac{p\left(D|\underline{\alpha},\underline{d},\eta\right)p\left(\underline{\alpha}|\underline{d},\eta\right)}
{p\left(D|\underline{d},\eta\right)}\log\left[\frac{p\left(D|\alpha,\underline{d},\eta\right)p\left(\alpha|\underline{d},\eta\right)}{p\left(\underline{\alpha}|\underline{d},\eta\right)p\left(D|\underline{d},\eta\right)}\right]\nonumber\\
&=&\int\!dD\;\int\!d\underline{\alpha}\;p\left(D|\underline{\alpha},\underline{d},\eta\right)p\left(\underline{\alpha}|\underline{d},\eta\right)\log\left[\frac{p\left(D|\underline{\alpha},\underline{d},\eta\right)}{p\left(D|\underline{d},\eta\right)}\right]\nonumber\\
&=&\int\!dD\;\int\!d\underline{\alpha}\;p\left(D|\underline{\alpha},\underline{d},\eta\right)p\left(\underline{\alpha}|\underline{d},\right)\log\left[\frac{p\left(D|\underline{\alpha},\underline{d},\eta\right)}
{\int\!d\underline{\alpha}\;p\left(D|\underline{\alpha},\underline{d},\eta\right)p\left(\underline{\alpha}|\underline{d}\right)}\right]
\label{eqn_expected_utility}
\end{eqnarray}
where we dropped the $\eta-$dependence of the posterior of $\underline{\alpha}$ in the last line, since our knowledge about $\underline{\alpha}$ is not influenced by a possible future
action.
Closer inspection of Eq. \ref{eqn_expected_utility} reveals that only two different probability distributions are required to compute the expected utility: the posterior distribution of
$\underline{\alpha}$ given the old data $\underline{d}$, $p\left(\underline{\alpha}|\underline{d}\right)$
and the likelihood of the new datum $D$ based on the previous measurements, $p\left(D|\underline{\alpha},\underline{d},\eta\right)$.
\subsection{The Linear Design}
Assuming that the concentration profile $c(x)$ depends linearly on the concentrations $c_{i}\left(x_{i}\right), i=1..q$
at a given set of $q$ support points $\underline{x}$ then Eq.\ref{equation_1}
can be recast in the following form
\begin{equation}
\underline{d}=\underline{f}+\underline{\epsilon}=\underline{\underline{M}}\,\underline{c}+\underline{\epsilon},
\end{equation}
where the data vector $\underline{d}$ is of size $p$, the matrix $\underline{\underline{M}}$ is a $p\mathrm{x}q-$matrix and the parameter-vector 
$\underline{c}$ has $q$ components.
However, the requirement of linearity applies only to the concentration parameter vector ${\underline{c}}$, the functional form of the concentration may be much more complex, e.g.  $c(x)=c_{1}*\left(x-x_{3}\right)^{4}+c_{2}*\sqrt{|x-x_{1}|}$, although almost always $c(x)$ is chosen to be constant between the different support points: $c(x)=c_{i}, \forall x\in \left[x_{i},x_{i+1}\right]$ or as linear interpolation between the support points.
The noise vector $\underline{\epsilon}$ is normally distributed $\underline{\epsilon}\sim{\phantom a} N\left(0,\underline{\underline{\Sigma^{-1}}}\right)$, where $\underline{\underline{\Sigma}}$ is a diagonal matrix with the entries $\underline{\underline{\Sigma}}_{ii}=1/\sigma_{i}^{2}$.
Every row of $\underline{\underline{M}}_{j}$ is given by the solution of Eq. \ref{equation_1} for a specified initial energy $E_{j}^{0}$, ${\underline{m}\left(E_{j}^{0}\right)}^{T}$. The consideration of the uncertainties in the entries of the matrix due to energy straggling of the impinging particles is beyond the scope of the present paper, but see e.g. \cite{Mayer08}.

With a Gaussian likelihood for the existing data and a flat prior for the parameters the posterior of the concentration vector reads
\begin{eqnarray}
p\left(\underline{c}|\underline{d},\eta\right)\propto p\left(\underline{d}|\underline{c},\eta\right)&=&\frac{1}{Z}\exp\left(-\frac{1}{2}\left(\underline{d}-\underline{\underline{M}}\,\underline{c}\right)^{T}\underline{\underline{\Sigma}}\left(\underline{d}-\underline{\underline{M}}\,\underline{c}\right)\right)\nonumber\\
&=&\sqrt{\frac{\det{\underline{\underline{A}}}}{\left(2\pi\right)^{q}}}\exp\left(-\frac{1}{2}\left(\underline{c}-\underline{c_{0}}\right)^{T}\underline{\underline{A}}\left(\underline{c}-\underline{c_{0}}\right)\right)
\end{eqnarray}
with
\begin{equation}
\underline{\underline{A}}={\underline{\underline{M}}}^{T}\underline{\underline{\Sigma}}\,\underline{\underline{M}}\;\;\;\;\mathrm{and}\;\;\;\;\underline{c_{0}}={\underline{\underline{A}}}^{-1}{\underline{\underline{M}}}^{T}\underline{\underline{\Sigma}}\,\underline{d}.
\end{equation}
The posterior distribution of $\underline{c}$ including the new data point $D$ with its uncertainty $\sigma$, $p\left(\underline{c}|D,\underline{d},\eta\right)$ can similarly be cast in a Gaussian form. Therefore Eq. \ref{eqn_expected_utility} can be solved analytically \cite{Fedorov72} and yields a simple closed form for the
exponential utility \cite{Dreier07,Preuss08}:
\begin{equation}
\mathrm{EU}\left(\underline{d},\eta\right)=\frac{1}{2}\left(\log\left(1+\mathrm{G}\right)-r\right)
\label{eq_EU_G}
\end{equation}
with
\begin{equation}
G=\frac{\underline{m}\left(\eta\right)^{T}\underline{\underline{A}}^{-1}\underline{m}\left(\eta\right)}{\sigma^{2}}.
\label{eq_G}
\end{equation}
If $p\left(D|\underline{c},\eta\right)$ is Gaussian then $r=0$.
The variation of the EU depends on the vector $\underline{m}\left(\eta\right)$
which in turn is uniquely determined
by the choice of the initial energy $E_{p+1}^{0}$. The optimum (maximum of the EU) can be found by a simple 1-D scan of the energy.

\begin{figure}
\includegraphics*[width = 8.5 cm]{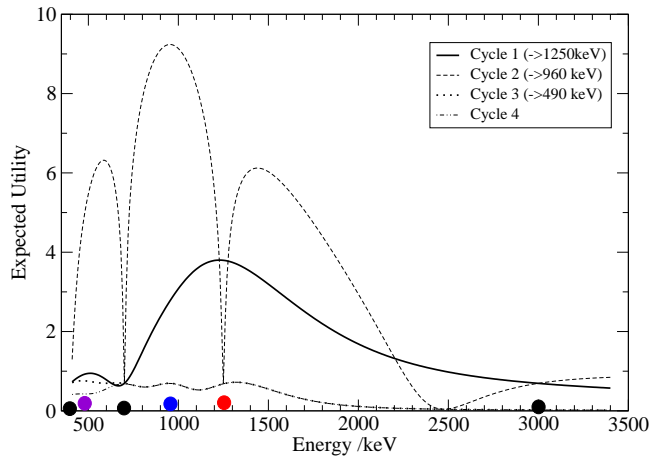}
\caption{\it Expected Utility for subsequent measurements. All measurements are performed with the initial energy $E^{0}$ suggested by the maximum of the EU in the corresponding cycle (indicated by marks on the energy axis).}
\label{fig_linearEU}
\end{figure}

The sequential design approach in action is displayed in Fig \ref{fig_linearEU}.
Starting from the surface the concentration at increasingly larger depth intervals is of interest. For this example the chosen depths are 0 nm, 80 nm, 240 nm, 470 nm and 950 nm.
After initial measurements at 400 keV, 700 keV and 3000 keV (representing the lower and upper limit of the useful energy range for the measurements and one calibration measurement) the best energy for the next measurement has to be determined.
The EU for this first cycle has a maximum at 1250keV (solid line).
After a measurement with this energy the EU for the next measurement has its maximum at 960keV and about twice the EU than before.
This, on the first glance, surprising increase of the EU can be made transparent: With 5 unknowns and 5 (informative) measurements the solution space of this linear problem no longer covers a sub-manifold of the parameter space:
It 'collapses' and the volume of the 'occupied' parameter space starts to be determined by the measurement uncertainties. Therefore the
5-th measurement has a very high EU.
In the following cycle(s) the amplitude of the EU is much lower since
the subsequent measurements now gradually shrink the 'volume' of the parameter posterior distribution.
As long as the EU is above the intended threshold for new measurements (which depends on the addressed physical problem) further measurements are indicated.

How much better is the BED derived experiment compared to an experiment with the same number but equidistant chosen initial energies? The entropy of the parameter posterior distribution would be the obvious quantity to compare. However, for the time being, many scientists are not happy with this measure and prefer a more familiar measure, e.g. the condition number.
The condition number of the (pseudo-)inverse of $\underline{\underline{M}}$ is often used to characterize linear least squares problems \cite{numrecipes} and is a measure how strongly uncertainties in the data vector $\underline{d}$ may be amplified by
multiplication with the pseudo-inverse matrix. Using this measure the BED optimized setting surpasses
the equidistant experiment by a factor of more than 100 (!).

\subsection{Non-linear Design}
The analytical solution in the preceeding case was possible because several approximations
have been applied: The integration range of the integration over the predicted datum (a positive quantity) had to be changed from
$\int_{0}^{\infty}\!dD$ to $\int_{-\infty}^{\infty}\!dD$. Given the actual number of counts and the uncertainties this can easily be justified.
Unfortunately, a similar change of the integration limits had to be applied also in the parameter integration
(from $\int_{0}^{1}\!d\underline{c}$ to $\int_{-\infty}^{+\infty}\!d\underline{c}$) and here it definitely affects the results.
The analysis could be repeated substituting the analytical integration by the numerical counterparts
(e.g. using codes like VEGAS \cite{numrecipes} or MCMC approaches).
Furthermore, the uncertainty of the predicted datum D is not constant but proportional to the signal $\sigma_{D}\propto D$
and therefore also the integrations over the data space have to be done numerically.
Under those circumstances there is no difference in the computation to a non-linear experimental design problem.
{\it Additionally} it turned out that the actual quantity of interest is the decay length of the hydrogen depth profile
and that quite accurate data for the surface hydrogen concentration are available (additionally measuring the ${}^{4}\mathrm{He}$ of the
$\mathrm{D}\left({}^{3}\mathrm{He},p\right){}^{4}\mathrm{He}$ reaction).
Therefore the optimal energy settings
for the estimation of the parameters $a_{1}$ and $a_{2}$ of concentration profiles of the functional form of Eq. \ref{eq_mock_data}
have to be computed. However, in non-linear experimental design the measured data influences the EU (in contrast to the linear case:
the maximum of the EU is independent of the actually measured data, cf. Eq. \ref{eq_EU_G}) and this poses a practical problem:
The next accelerator energy has to be determined after the previous measurement. And longer computation times to optimize the EU, causing delays, are not
tolerable.

Here the posterior sampling approach, suggested in \cite{Loredo03}, proved very valuable. It turned out that sets
of posterior samples $p\left(a_{1i},a_{2i}|\underline{d},\eta\right),i=1..N$ drawn from $p\left(a_{1},a_{2}|\underline{d},\eta\right)$
could be generated quite efficiently (partly due to the low dimensionality of the parameter space). With that sample (typically of size 1000) the denominator of the logarithm in Eq. \ref{eqn_expected_utility}
is given by a simple summation
\begin{equation}
\int\!d\underline{\alpha}\,p\left(D|\underline{\alpha},\underline{d},\eta\right)p\left(\underline{\alpha}|\underline{d}\right)\approx\sum_{i=1}^{N}p\left(D|\underline{\alpha_{i}},\underline{d},\eta\right).
\end{equation}
The biggest saving comes from the fact that the posterior sample is independent from the actual value of $D$ and of the design action $\eta$:
all computations are reduced to repeated evaluations of the likelihood, which can efficiently be vectorized. Finding the best energy is a matter of
less than 5 minutes(!) on contemporary hardware (Linux-PC, 2GHz).

\begin{figure}
\includegraphics*[width=14cm]{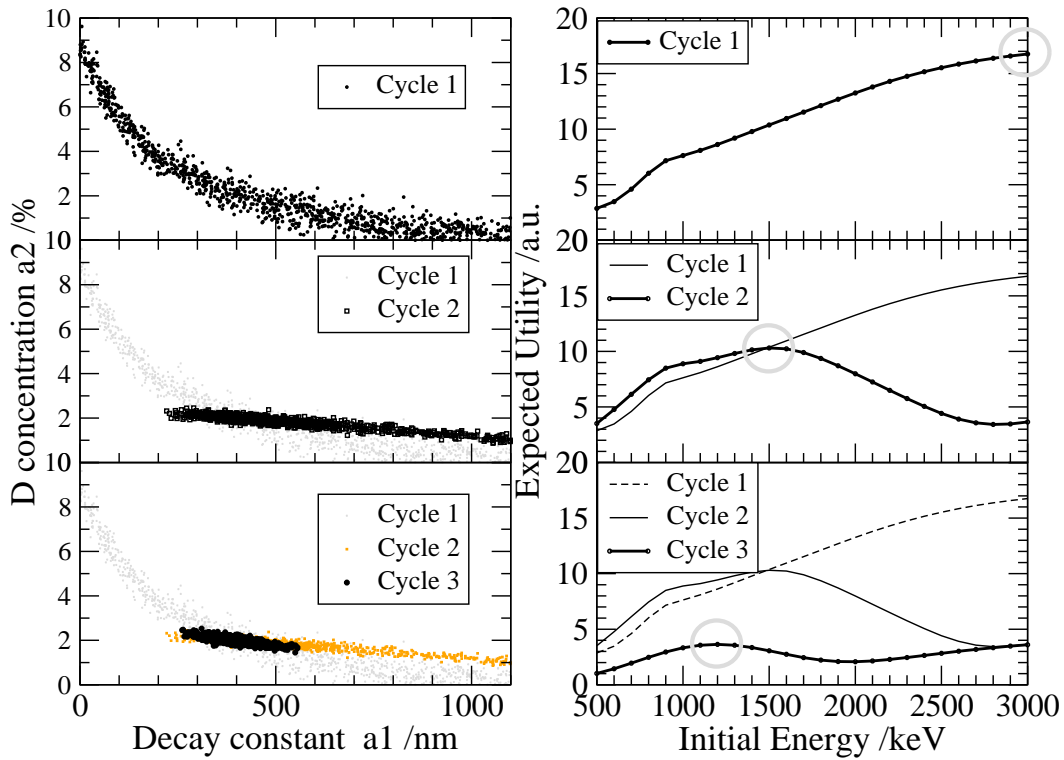}
\caption{\it Three cycles of the Experimental Design process: On the left hand side 1000 samples drawn from the posterior distribution $p\left(a_{1},a_{2}|\underline{d},\eta\right)$
are displayed.
On the right hand side the EU is plotted and the maximum is indicated by a circle. The corresponding abscissa value is the suggested next measurement energy. Performingthat
measurement yields the posterior distribution given in the next row.}
\label{fig_nonlinearEU}
\end{figure}

In Fig. \ref{fig_nonlinearEU} three cycles of the non-linear BED are shown: After a first measurement at 500keV the posterior distribution
of $\left\{a_{1},a_{2}\right\}$ is visualized in the upper left graph by the posterior sample. The single measurement does not allow to distinguish between
a large decay constant $a_{1}$ and low constant offset $a_{2}$ or vice versa. The EU, plotted in the upper right graph,
favors now a measurement at the other end of the energy range (the maximum of the utility function
is encircled). After a measurement with 3MeV ${}^{3}\mathrm{He}$ the 'area' of the posterior distribution is
significantly reduced (middle row, left graph): The background concentration is below 3\% but the decay length is still quite undetermined.
The EU has a maximum at 1500 keV, still with a pretty high EU. Performing a measurement with 1500keV localizes the posterior distribution
around the true (but unknown value of $a_{1}=395$nm and $a_{2}=0.02$). The next measurement should be performed at 1200keV but the EU is significantly lower
than before: subsequent measurements are predominantly improving the statistics: a second measurement at 3 MeV provides nearly the same information.

\section{Conclusion and Outlook}
The concept of Bayesian Experimental Design allows to objectively optimize experimental designs. Here we presented two different approaches to optimize
NRA depth profiling: First in a linear setting, allowing an analytical solution and straightforward parametric studies. Second, a time-critical non-linear
experimental design problem which could be tackled using posterior sampling. Both optimization procedures may considerably increase the accuracy of the
derived depth profiles compared to the present approach and at the same time reduce the overall measurement time by signaling a diminishing utility of further
measurements.
With the posterior sampling approach many sequential measurements can now be optimized on the fly: This opens up the door for a wealth of new applications of BED in
the field of ion beam analysis\cite{Toussaint2000} as well as in other physical areas \cite{Loredo03,Preuss08,Toussaint2006}


\begin{thebibliography}{99}
\bibitem{Iter08} ITER Organization, http://www.iter.org, (2008).

\bibitem{Toussaint2000} U. von~Toussaint and R. Fischer and K. Krieger and V. Dose,
Depth Profile Determination with Confidence Intervals from Rutherford Backscattering Data,
New Journal of Physics \textbf{1}, 11 (1999).

\bibitem{Alimov05} V. Kh. Alimov and M. Mayer and J. Roth,
Differential cross-section of the D$\left({}^{3}\mathrm{He},p\right){}^{4}\mathrm{He}$ nuclear reaction and depth
profiling of deuterium up to large depths, Nucl. Instr. Meth. B \textbf{234}, 169-175 (2005).

\bibitem{Moeller80} W. M{\"o}ller and F. Besenbacher, A note on the ${}^{3}$He+D nuclear reaction cross section, Nucl. Instr. and Meth. \textbf{168}(1), 111-114 (1980).

\bibitem{Bosch92}H.-S. Bosch and G. M. Hale, Improved formulas for fusion cross-sections and thermal reactivities, Nucl. Fusion \textbf{32}, 611-632 (1992).

\bibitem{Alimov08} V. Kh. Alimov et al,
Deuterium retention in tungsten exposed to low-energy, high-flux clean and carbon-seeded deuterium plasmas,
J. Nucl. Mat. \textbf{375}, 192-201 (2008).

\bibitem{Tesmer95} J. R. Tesmer and M. Nastasi and J.C. Barbour and C. J. Maggiore and J. W. Mayer,
Handbook of Modern Ion Beam Analysis, Materials Research Society, Pittsburgh, PA, USA (1995).

\bibitem{Clyde95} M. Clyde and P. M{\"u}ller and G. Parmigiani,
Exploring expected utility surfaces by markov chains, Source: http://ftp.stat.duke.edu/WorkingPapers/95-39.ps (1995).

\bibitem{MacKay92} D. MacKay,
Information-based objective functions for active data selection, Neural Computation \textbf{4}(4), 590-604 (1992).

\bibitem{Lindley56} D. V. Lindley,
On the measure of information provided by an experiment, Ann. Stat \textbf{27}, 986-1005 (1956).

\bibitem{Fedorov72} V. V. Fedorov,
Theory of Optimal Experiments, Academic, New York (1972).

\bibitem{Loredo03} T. J. Loredo,
'Bayesian Adaptive Exploration' in {\it Bayesian Inference and Maximum Entropy Methods in Science and Engineering},
edited by G. Erickson and Y. Zhai, AIP, Melville, NY, vol. Conf. Proc \textbf{707}, 330-346 (2003).

\bibitem{Fischer04} R. Fischer,
'Bayesian Experimental Design - Studies for Fusion Diagnostics' in {\it Bayesian Inference and Maximum Entropy Methods in Science and Engineering}, edited by R. Fischer, R. Preuss and U. von Toussaint, AIP, Melville, NY, vol. Conf. Proc \textbf{735}, 76-83 (2004).

\bibitem{Caticha00}P. Riegler and N. Caticha,
'MaxEnt queries and sequential sampling' in {\it Bayesian Inference and Maximum Entropy Methods in Science and Engineering}, edited by A. Mohammad-Djafari, AIP, Melville, vol. Conf. Proc \textbf{568}, 270-279 (2001).

\bibitem{Dreier07} H. Dreier,
Bayesian Experimental Design: Applications in Nuclear Fusion, PhD-thesis, IPP-Report 13/8, Max-Planck-Institut f{\"u}r Plasmaphysik (2007).

\bibitem{Preuss08} R. Preuss and H. Dreier and A. Dinklage and V. Dose,
Data adaptive control parameter estimation for scaling laws for magnetic fusion devices,
EPL \textbf{81}(5), 55001 (2008).

\bibitem{Chaloner95} K. Chaloner and I. Verdinelli,
Bayesian experimental design: A review, Stat. Sci. \textbf{10}, 273-304 (1995).

\bibitem{Bernardo79} J. M. Bernardo,
Expected Information as Expected Utility, Ann. Stat. \textbf{7}, 686-690 (1979).

\bibitem{Mayer08} M. Mayer, E. Gauthier, K. Sugiyama, and U. von Toussaint,
Quantitative depth profiling of Deuterium up to very large depths, to be submitted.

\bibitem{numrecipes} W. H. Press and S. A. Teukolsky and W. T. Vetterling and B. P. Flannery,
Numerical Recipes in Fortran 77, Oxford Science Publications, Cambridge University Press, 2nd edition (1992).

\bibitem{Toussaint2006} U. von~Toussaint and V. Dose,
Bayesian Analysis in surface physics,
Applied Physics A \textbf{82}, 403-413  (2006).

\end{thebibliography}
\end{document}